\documentclass[apj,numberedappendix]{emulateapj}
\usepackage{apjfonts,longtable}
\usepackage{ulem}

\newcommand{\be}{\begin{eqnarray}}
\newcommand{\ee}{\end{eqnarray}}

\newcommand{\lp}{\left(}
\newcommand{\rp}{\right)}
\newcommand{\lb}{\left[}
\newcommand{\rb}{\right]}

\newcommand{\emin}{E_{\rm min}}
\newcommand{\emax}{E_{\rm max}}

\begin{document}

\slugcomment{Submitted for publication in The Astrophysical Journal Letters}
\shorttitle{Radio Transients from AIC}
\shortauthors{Piro, A. L. \& Kulkarni, S. R.}

\normalsize


\title{Radio Transients from the Accretion Induced Collapse of White Dwarfs}

\author{Anthony L. Piro and S.  R. Kulkarni}

\affil{Cahill Center for Astrophysics, California Institute of Technology, Pasadena, CA 91125, USA; piro@caltech.edu}


\begin{abstract}
It has long been expected that in some scenarios when a white dwarf (WD) grows to the Chandrasekhar limit, it can undergo an accretion induced collapse (AIC) to form a rapidly rotating neutron star. Nevertheless, the detection of such events has so far evaded discovery, likely because the optical, supernova-like emission is expected to be dim and short-lived.  Here we propose a novel signature of AIC: a transient radio source lasting for a few months. Rapid rotation along with flux freezing and dynamo action can grow the WD's magnetic field to magnetar strengths during collapse. The spindown of this newly born magnetar generates a pulsar wind nebula (PWN) within the \mbox{$\sim10^{-3}-10^{-1}\,M_\odot$} of ejecta surrounding it. Our calculations show that synchrotron emission from the PWN may be detectable in the radio, even if the magnetar has a rather modest magnetic field of $\sim2\times10^{14}\,{\rm G}$ and an initial spin period of $\sim10\,{\rm ms}$. An all-sky survey with a detection limit of $1\ {\rm mJy}$ at $1.4\ {\rm GHz}$ would see $\sim4(f/10^{-2})$ above threshold at any given time, where $f$ is the ratio of the AIC rate to Type Ia supernova rate. A similar scenario may result from binary neutron stars if some mergers produce massive neutron stars rather than black holes. We conclude with a discussion of the detectability of these types of radio sources in an era of facilities with high mapping speeds. 
\end{abstract}



\section{Introduction}
\label{sec:Introduction}

As an accreting white dwarf (WD) grows toward the Chandrasekhar
limit, a well-known potential outcome is ignition of its nuclear
fuel, leading to a Type Ia supernova \citep[SN Ia,][]{HillebrandtNiemeyer2000}.
However in some cases (e.g. mass transfer onto O/Ne/Mg
WDs and C/O WD mergers; \citealp{CanalSchatzman1976,NomotoKondo1991}) electron capture can
rob the core of its degeneracy pressure support  leading to formation
of a neutron star (NS). This ``Accretion Induced Collapse'' (AIC) has
been invoked to explain millisecond pulsars \citep[e.g.][]{Bhattacharya1991}, subsets of gamma-ray bursts \citep[e.g.][]{Daretal1992,Metzgeretal2008b}, magnetars \citep[e.g.][]{Usov1992}, and may be a source
of \mbox{$r$-process} nucleosynthesis \citep{Hartmannetal1985,Fryeretal1999}.

Despite its potential importance, there has been no reported detection of an
AIC event. To start with, the expected AIC rate is no more than
$\approx1\%$ of that of SNe Ia \citep{YungelsonLivio1998}. Next, relative to
Type I and Type II SNe, the ejecta mass is expected to be
small ($\lesssim10^{-1}\,M_\odot$), produce little $^{56}$Ni ($\lesssim10^{-2}\,M_\odot$),
and move at high velocity ($\approx0.1c$). The resulting
optical transient is thus considerably fainter
than a typical SN (5 magnitudes or more) and lasts \mbox{$\sim1\,{\rm day}$}
\citep{Metzgeretal2009b,Darbhaetal2010}.

Rapid rotation should accompany AIC due to the accretion
of mass and angular momentum. Furthermore, a strong magnetic field
may be amplified through flux freezing during collapse and via dynamo
action \citep{DuncanThompson1992}. Therefore a plausible
outcome of AIC is creation of a quickly spinning magnetar
\citep{Usov1992,Kingetal2001,Levanetal2006}.  The WD collapse unbinds
material \citep{Dessartetal2006}, and the remnant disk loses mass via outflows
driven by neutrino heating, turbulent viscosity, and recombination
of free nuclei into helium
\citep{LeeRamirezRuiz2007,Metzgeretal2008a,Metzgeretal2009a,Leeetal2009}. This leads to
$M_{\rm ej}\approx10^{-3}-10^{-1}\,M_\odot$ of ejecta with
velocity $v_{\rm ej}\approx0.1c$.
Such a configuration will also follow a double NS merger
if the total binary mass is below the maximum NS mass. 

Here we investigate the detectability of the pulsar wind nebula (PWN)
powered by a newly formed magnetar following AIC. In \S \ref{sec:dynamics} we describe a
model for how the PWN expands into the surrounding ejecta. In \S \ref{sec:radio} we estimate the radio spectrum and lightcurve, showing that the PWN is a transient radio source for a few months. We estimate the detection rate in \S \ref{sec:rates} and discuss the detectability of
such events with soon-to-be-commissioned  high speed radio mapping machines in \S \ref{sec:DetectionIdentification}.

\section{Dynamics of the Pulsar Wind Nebula}
\label{sec:dynamics}

Following the implosion of the WD and subsequent disk outflows, the ejecta expands with velocity $v_{\rm ej}\approx0.1c$ and kinetic energy $E_{\rm ej}\approx M_{\rm ej}v_{\rm ej}^2/2\approx10^{50}\,{\rm erg}$. This plows into the ISM with particle density $n_0$, which can vary greatly from $\sim~10^{-6}\,{\rm cm^{-3}}$ for events outside of their host galaxy to $\sim~1\,{\rm cm^{-3}}$ in denser environments. Once the ejecta has swept up a mass comparable to its own, it decelerate as it enters the Sedov-Taylor phase on a timescale \citep{McKeeTruelove1995}
\be
	\tau_{\rm ST} = 0.5E_{\rm ej}^{-1/2}M_{\rm ej}^{5/6}(m_pn_0)^{-1/3}\approx16 E_{50}^{-1/2}M_{-2}^{5/6} n_0^{-1/2}{\rm yr},
	\nonumber
	\\
\ee
where $E_{50}=E_{\rm ej}/10^{50}\,{\rm erg}$ and $M_{-2}=M_{\rm ej}/10^{-2}\,M_\odot$.
Since $\tau_{\rm ST}$ is much longer than the times of interest (especially when $n_0$ is small), the ejecta is always in an ejecta-dominated (or free-expansion) phase. In this case the ejecta maintains constant velocity with radius $R_{\rm ej}\approx v_{\rm ej}t$. The forward shock is merely a distance $\approx(4/3)^{1/3}R_{\rm ej}$ ahead of $R_{\rm ej}$, and the reverse shock has barely developed behind it. The pressure behind the forward shock and down to the reverse shock is roughly constant and given by the strong shock limit $\approx(4/3) m_pn_0v_{\rm ej}^2$.

The magnetar injects energy in the form of magnetic fields and relativistic particles at a rate
\be
	L(t) = L_0/(1+t/\tau)^p.
\ee
Here we assume dipole spindown\footnote{When the magnetar is spinning $P_0\lesssim 3\,{\rm ms}$ and readily radiating neutrinos, this can give rise to a neutrino-driven, magneto-centrifugal wind \citep{Thompsonetal2004} which could greatly enhance the spindown.} and thus $p=2$. For a magnetic moment $\mu$, initial spin frequency $\Omega_0=2\pi/P_0$, and moment of inertia $I=0.35M_*R_*^2$ \citep{LattimerPrakash2001},
\be
	L_0 = \mu^2\Omega_0^4/6c^3 = 1.2\times10^{47}\mu_{33}^2P_3^{-4}\,{\rm erg\ s^{-1}},
\ee
and
\be
	\tau = 6Ic^3/\mu^2\Omega_0^2 = 4.8\times10^4\mu_{33}^{-2}P_3^2\,{\rm s},
\ee
where $\mu_{33}=\mu/10^{33}\,{\rm G\ cm^{3}}$, $P_3=P_0/3\,{\rm ms}$, and we use $M_*=1.4\,M_\odot$ and $R_*=12\,{\rm km}$.

The energy input from the magnetar powers a PWN with radius $R_p$.
Deeper inside, a wind termination shock forms at radius $R_t$, where the ram pressure of the pulsar wind equals the PWN pressure $P$ \citep{GaenslerSlane2006},
\be
	R_t \approx (L/4\pi c P)^{1/2},
\ee
where $R_t\ll R_p$. The general picture developed from galactic PWNe is that particles are accelerated at or near $R_t$. This seeds the PWN with relativistic electrons out to $R_p$. 

A central issue is the content of the PWN. As a pulsar wind flows from the light cylinder, it is inferred to have a large magnetization \citep[$\sigma\sim10^4$, where $\sigma$ is the ratio of Poynting flux to particle energy flux;][]{Arons2002}. However multiple lines of evidence, including the expansion velocities and high-energy modeling of PWNe, require that the magnetization decreases significantly \mbox{($\sigma\ll1$)} by the time the wind reaches $R_t$. In the region where the 
radiation arises, $R_t < r< R_p$, we define $\eta_e$ and $\eta_B$ as the fraction of the magnetar luminosity that goes into electrons and magnetic fields, respectively, with typical values of $\eta_e\approx0.999$ and $\eta_B\approx10^{-3}$.

Following \citet{ReynoldsChevalier1984} we assume the energy density of electrons and magnetic fields evolve independently and obey a relativistic equation of state. Thus 
\be
	\frac{d}{dt}(4\pi P_eR_p^4) = (\eta_eL-\Lambda)R_p,
	\label{eq:energy_e}
\ee
where $\Lambda$ is the radiative loss rate, and
\be
	\frac{d}{dt}(4\pi P_BR_p^4) = \eta_B LR_p,
	\label{eq:energy_b}
\ee
where $P_B=B^2/8\pi$. Note that in the adiabatic limit $\eta_BL\approx0$, this predicts $P_B\propto R_p^{-4}$ and $B\propto R_p^{-2}$, as expected from flux freezing. Momentum conservation is given by
\be
	M_s\frac{d^2R_p}{dt^2} = 4\pi R_p^2 \lb P - \rho_{\rm ej}\lp v_p - \frac{R_p}{t} \rp^2 \rb,
	\label{eq:momentum}
\ee
where $v_p=dR_p/dt$ and $P=P_e+P_B$ is the total pressure.

The PWN sweeps up ejecta and creates a shell with mass
\be
	M_s = \left\{ 
	\begin{array}{l l}
  M_{\rm ej}(R_p/v_{\rm ej}t)^3, &\ R_p<v_{\rm ej}t\\
  M_{\rm ej},  &\ R_p\ge v_{\rm ej}t\\ \end{array}.		\right.
  \label{eq:ms}
\ee
Once $R_p\approx v_{\rm ej}t$, the PWN reaches material that has been shock-heated by its interaction with the ISM. Since we generally find $P\gg (4/3)m_pn_0v_{\rm ej}^2$, we do not expect the shock-heated pressure to have a large dynamical effect on the PWN. This is in contrast to PWNe growing within a SN remnant, where the large pressure behind the reverse shock (on a timescale $t\gg\tau_{\rm ST}$) can cause significant compression and magnetic field amplification \citep{ReynoldsChevalier1984}.

Equations (\ref{eq:energy_e}),  (\ref{eq:energy_b}), and (\ref{eq:momentum}) with $dR_p/dt = v_p$ provide four first-order differential equations for the four dependent variables $P_e$, $P_B$, $R_p$, and $v_p$, respectively. For simplicity, we calculate the PWN evolution using $\Lambda\approx0$, so that equations~(\ref{eq:energy_e}) and  (\ref{eq:energy_b}) are combined using $P=P_e+P_B$ and $\eta_e+\eta_B=1$. When $L$ is constant, the analytic result is \citep{Chevalier1977}
\be
	R_p \approx
	1.3\times10^{16}
	L_{47}^{1/5}
	E_{50}^{3/10}
	M_{-2}^{-1/2}
	t_6^{6/5}\,{\rm cm},
	\label{eq:r}
\ee
and
\be
	B \approx
	 6\eta_{B,-3}^{1/5}
	L_{47}^{1/5}
	E_{50}^{-9/20}
	M_{-2}^{3/4}
	t_6^{-13/10}\,{\rm G},
	\label{eq:b}
\ee
where $L_{47}=L/10^{47}\,{\rm erg\ s^{-1}}$, $t_6=t/10^6\,{\rm s}$, and $\eta_{B,-3}=\eta_B/10^{-3}$. This gives some idea of the rough values expected, although for our detailed calculations, $R_p$ and $B$ deviate slightly from these scalings when $t\gtrsim\tau$.

Rayleigh-Taylor instabilities can act at $R_p$ due to the low density PWN pushing up against high density ejecta. The growth time for a large density contrast is $\tau_{\rm RT} \approx (g_{\rm eff}k)^{-1/2}$, where $g_{\rm eff}$ is the effective gravitational acceleration at the boundary and $k$ the wavenumber. For $R_p$ given by equation~(\ref{eq:r}), $g_{\rm eff}\approx d^2R_p/dt^2\approx (6/25)R_p/t^2$ and the growth rate for $k\approx n/R_p$ is $\tau_{\rm RT}\approx (25/6n)^{1/2}t$. For sufficiently small wavelengths (large $n$), $\tau_{\rm RT}\lesssim t$ and instability results. This is not surprising since similar systems, like the Crab Nebula, have morphologies strongly impacted by instabilities. But for the current analysis we do not include this complication.

\section{Synchrotron Radio Emission}
\label{sec:radio}

A power-law spectrum of relativistic electrons $n(E)=K E^{-s}$ are accelerated near the termination shock at $R_t$ and fill the PWN out to $R_p$, where $n(E)$ is in units of ${\rm electrons\ erg^{-1}\ cm^{-3}}$. The power-law parameters within the PWN can change via cooling and injection of new electrons \citep[e.g.][]{Gelfandetal2009,Bucciantinietal2011}. For the present work we estimate the synchrotron spectrum at any time by  fixing $P_e$ and $P_B$ from our dynamical calculations.

Our discussion of synchrotron emission largely follows the work of \citet{Pacholczyk1970}. Synchrotron emission is 
self absorbed below a frequency
\be
	\nu_{\rm SA} = 2c_1(R_pc_6)^{2/(s+4)}K^{2/(s+4)}(B\sin\theta)^{(s+2)/(s+4)},
	\label{eq:nusa}
\ee
where $c_1=6.27\times10^{18}$ in cgs units, $c_6$ depends on $s$ and can be found in Appendix 2 of \citet{Pacholczyk1970}, and $\theta$ is the pitch angle. Throughout we assume an average of $\sin\theta=(2/3)^{1/2}$. In the optically thick limit the flux is
\be
	F_{\nu} = \frac{\pi R_p^2}{D^2}\frac{c_5}{c_6}(B\sin\theta)^{-1/2}\lp\frac{\nu}{2c_1} \rp^{5/2},
\ee
where $D$ is the distance and $c_5$ is another constant. In the optically thin limit,
\be
	F_{\nu} = \frac{4\pi R_p^3}{3D^2}c_5K(B\sin\theta)^{(s+1)/2}\lp\frac{\nu}{2c_1} \rp^{-(s-1)/2},
	\label{eq:fnu2}
\ee
which assumes an emission filling factor of order unity. We use a simple interpolation between these two limits for the total emission spectrum $F_\nu$. The location of the spectrum's peak at $\nu_{\rm SA}\propto B^{(s+2)/(s+4)}$ shifts to lower frequencies as $B$ decreases during expansion (eq. [\ref{eq:b}]). The peak flux scales $\propto B^{(2s+3)/(s+4)}$, and thus also decreases with time.
 
The synchrotron cooling time scales as $E/|\dot{E}|\propto \nu^{-1/2}$. Therefore there is a $\nu_c$ above which synchrotron cooling beats adiabatic expansion \citep{ReynoldsChevalier1984}
\be
	\nu_c = \frac{c_1}{c_2^2B^3\sin^3\theta}\lp \frac{v_p}{R_p}\rp^2,
	\label{eq:nuad}
\ee
where $c_2= 2.37\times10^{-3}$ in cgs units. Electrons that evolve adiabatically maintain their spectrum, while electrons that cool from synchrotron emission steepen, therefore
\be
	n(E) = \left\{ 
	\begin{array}{l l}
  KE^{-s}, &\ E<E_c\\
  KE_cE^{-(s+1)},  &\ E\ge E_c\\ \end{array},		\right.
\ee
where $E_c$ is the energy of electrons emitting a frequency $\nu_c$.
The prefactor $K$ is set by the total energy density of electrons $U_e=3P_e$, found by integrating
\be
	U_e &=& \int_{\emin}^{\emax}n(E)EdE,
	\label{eq:normalization}
\ee
where $E_{\rm min}$ and $E_{\rm max}$ are the minimum and maximum energies of electron spectrum, respectively.
For $s\approx1.5$ and the limit $\emax\gg E_c\gg \emin$, $U_e\approx 4KE_c^{2-s}$. The energy density is roughly set by $E_c$ since the spectrum steepens for $E>E_c$.

We next consider potential free-free absorption. The absorption coefficient is \citep{RybickiLightman1979}
\be
	\alpha_{\rm ff} \approx 1.9\times10^{-2}T^{-3/2}Z^2 n_e n_i \nu^{-2}g_{\rm ff}\,{\rm cm^{-1}},
	\label{eq:ff}
\ee
for $h\nu\ll k_{\rm B}T$, where $T$ is the temperature, $Z$ is the average charge per ion, $n_e$ and $n_i$ are the electron and ion densities, respectively, $g_{\rm ff}\sim1$ is the Gaunt factor, and all quantities are in cgs units.  Just behind the forward shocks
(see \S\ref{sec:dynamics}), the temperature is high ($\gg10^9\,{\rm K}$), and free-free absorption is negligible. But at the front edge of the ejecta, with density $\rho_{\rm ej}$ and temperature $T$, pressure continuity across the contact discontinuity requires $\rho_{\rm ej}k_{\rm B}T/m_p\sim n_0m_pv_s^2$. From this we estimate that $T\sim 10^5-10^8\,{\rm K}$ at $t\sim10^6-10^7\,{\rm s}$. For $Z/A\approx 1/2$ (where $A$ is the average atomic number) we estimate $Z^2n_en_i\sim (\rho_{\rm ej}/m_p)^2$. The observed flux is thus  $F_{\nu,\rm obs} \approx F_\nu e^{-\tau_{\rm ff}}$, where $\tau_{\rm ff} \approx \alpha_{\rm ff} \Delta R_p$ and $\Delta R_p\approx0.1R_p$ is the shell thickness.

 \begin{figure}
\epsscale{1.14}
\plotone{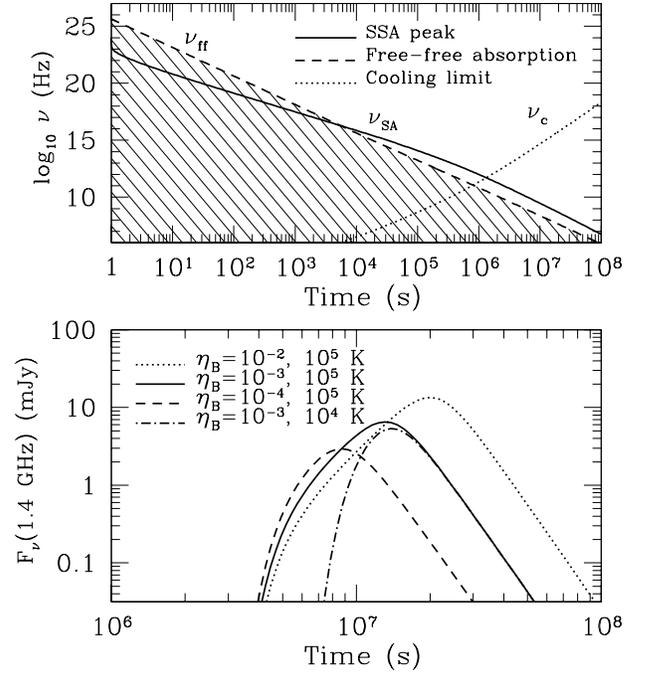}
\caption{The top panel shows the evolution of the critical frequencies with time for $M_{\rm ej}=10^{-2}\,M_\odot$, $v_{\rm ej}=0.1c$, $\mu=10^{33}\,{\rm G\ cm^3}$, and $P_0=3\,{\rm ms}$. The electron spectrum has a power-law index $s=1.5$. The shaded region shows where free-free absorption suppresses the synchrotron spectrum for $T=10^5\,{\rm K}$. This moves up for lower $T$ (see eq. [\ref{eq:ff}]). The spectrum peaks at frequency $\nu_{\rm SA}$. The bottom panel shows the time-dependent flux at $\nu=1.4\,{\rm GHz}$ for a distance $D=100\,{\rm Mpc}$ over a range of parameters as labeled.}
\label{fig:combine}
\epsscale{1.0}
\end{figure}

In the top panel of Figure \ref{fig:combine}, we plot the evolution of the key frequencies $\nu_{\rm SA}$ (eq. [\ref{eq:nusa}]),  $\nu_c$ (eq. [\ref{eq:nuad}]), and $\nu_{\rm ff}$, where the latter is defined by $\tau_{\rm ff}\approx 1$. From this one can follow the spectrum peak (at $\nu_{\rm SA}$) and determine when it is detectable. For example, $10\,{\rm GHz}$ emission peaks at $\approx6\times10^6\,{\rm s}$. In the bottom panel we explore the lightcurves at $1.4\,{\rm GHz}$ as we vary $T$ and $\eta_B$. Typical timescales are $\sim\,{\rm months}$ with a peak of $\sim3-10\,{\rm mJy}$. Free-free absorption is mostly negligible unless $T\lesssim10^4\,{\rm K}$. The peak flux and time of peak are sensitive to $\eta_B$.

\begin{figure}
\epsscale{1.14}
\plotone{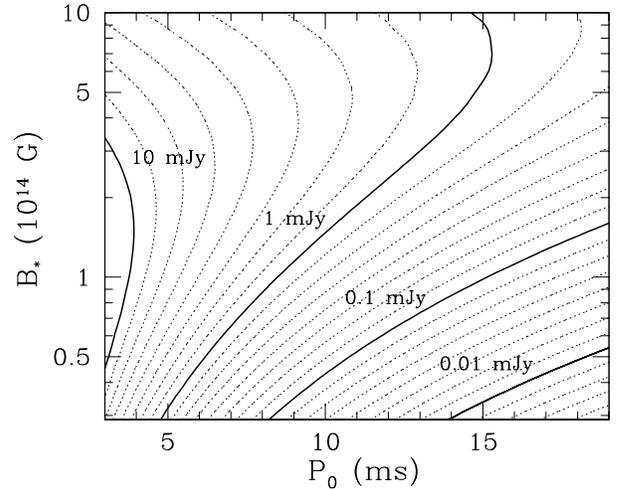}
\caption{The peak flux at $1.4\,{\rm GHz}$ at $100\,{\rm Mpc}$ as a function of the magnetar field $B_*$ and its initial period $P_0$. The contours of constant peak flux are spaced logarithmically in units of ${\rm mJy}$ with the thick, solid curves marking the labeled contours. In all cases we use $\eta_B=10^{-3}$, $T=10^5\,{\rm K}$, and $s=1.5$.}
\label{fig:contour_peak}
\epsscale{1.0}
\end{figure}

In Figure \ref{fig:contour_peak} we plot the peak flux at $1.4\,{\rm GHz}$ for a range of dipole field strengths and initial spin periods. The general trend is that faster spins and larger fields result in a more luminous radio source. This reverses in the top left corner where $\tau$ is sufficiently short that $L(t)$ has decreased significantly by the time of peak. Over most of this parameter space, the magnetar winds are not expected to be strong enough to generate a collimated outflow \citep{Bucciantinietal2012}.

\section{Detection Rates}
\label{sec:rates}

AIC can result from both channels popularly discussed for SNe Ia (single-degenerate and double-degenerate;
see \S\ref{sec:Introduction}). Thus it makes sense to normalize the AIC rate to that of the SN Ia rate.
The Lick Observatory Supernova Search finds a rate of $(3.01\pm0.062)\times10^{-5}\,{\rm Ia\ Mpc^{-3}\ yr^{-1}}$ \citep{Lietal2011}, which corresponds to $(4.0-7.1)\times10^{-3}\,{\rm Ia\  yr^{-1}}$ for the Milky Way.
Using population synthesis, \citet{YungelsonLivio1998} find AIC rates of
$8\times10^{-7}-8\times10^{-5}\,{\rm AIC\ yr^{-1}}$ for the Milky Way, depending on assumptions about the common-envelope phase and mass transfer. Their upper bound is similar to the constraint obtained from observed abundances of neutron rich
isotopes \citep{Hartmannetal1985,Fryeretal1999}.

For a typical peak $1.4\,{\rm GHz}$ flux of $F_{p,100}\sim5\,{\rm mJy}$ at a distance of $100\,{\rm Mpc}$,
and an AIC rate that is a fraction $f$ of the SN Ia rate, the detection rate of radio transients from AIC is
\be
	{\rm Rate} (>F_{p,100}) \approx 14
	\lp \frac{f}{10^{-2}} \rp
	\lp\frac{F_{p,100}}{5\,{\rm mJy}} \rp^{-3/2}
	{\rm yr^{-1}},
	\label{eq:detectionrate}
\ee
where $f\sim10^{-4}-10^{-2}$. 
For a \mbox{$\sim3\ {\rm month}$} duration emitting above \mbox{$\sim1\,{\rm mJy}$,} we expect $\sim4(f/10^{-2})$ AICs above threshold at a given time. In contrast, merely a few AICs are estimated to be detected as kilonovae per year \citep{Metzgeretal2009b}, and this number could be much less if a lack of differential rotation during WD accretion \citep{Piro2008} inhibits disk formation upon collapse \citep{Abdikamalovetal2010}

Similar radio emission is possible if some NS mergers produce a massive NS. This would then provide an electromagnetic counterpart following months after the gravitational wave emission during coalescence\footnote{Also see the work of \citet{NakarPiran2011}, which focuses on emission from the interaction of ejecta with the ISM.} \citep{MetzgerBerger2012,Nissankeetal2012}. In fact, the distance at which the radio emission can be detected is similar to that probed by the next generation of ``advanced'' ground-based laser-interferometers. The NS merger rate is comparable or greater than the AIC rate. For example, \citet{Kimetal2005} estimate a Galactic rate of $\sim(0.1-3)\times10^{-4}\,{\rm yr^{-1}}$. The detection frequency is thus similar to equation (\ref{eq:detectionrate}) with $f=10^{-2}$, with the caveat that it may be much less if most NS mergers produce black holes.

\section{Detection \&\ Identification of AIC Events}
\label{sec:DetectionIdentification}

This is a timely topic to discuss given the renaissance that is now occurring in decimetric radio astronomy.
Refurbished (EVLA) and new facilities promise high mapping speeds by using small diameter antennas to realize
a given total area (e.g. MeerKAT), through the use of focal plane arrays instead of a single feed to gain massive multiplex advantage 
(APERTIF), or both (ASKAP). 

We offer a plausible project aimed at detection of AICs that can be undertaken over the next few years. 
For a 7$\sigma$ detection\footnote{In using a low threshold (7$\sigma$) we take advantage of the fact that 
galaxies in the local universe (say out to 100\,Mpc) occupy a small solid angle ($\sim100$ square degrees,
excluding the nearest large galaxies), so chance coincidences
are suppressed by three orders of magnitude \citep{KulkarniKasliwal2009}.}
 of  a 1\,mJy point source, the EVLA mapping speed
in the 1.4\,GHz band (for a 500\,MHz band width)  is $\approx86\eta$ square degree per hour, where $\eta$ is the efficiency of time spent integrating on the sky (as opposed to slewing antennas or observing
calibrator sources). For ``on-the-fly'' mapping, $\eta=0.9$ (S. Myers, private communication) and a 250-hour allocation results in observing $\approx19,000$ square degrees
(comparable in area and depth to FIRST; \citealp{Beckeretal1995}).  
Another strategy would be to use the same allocation for several epochs focused on fields selected by their richness in nearby galaxies
\citep[as was done for the PTF key project  ``Investigation  of Transients in the Local Universe,''][]{Kasliwal2011}.
A systematic optical mapping of the sky should be undertaken a week or more prior to the radio observations and continued for a week after
(for instance with PTF, \citealp{Lawetal2009,Rauetal2009}). 

The EVLA radio survey would yield roughly two AICs. 
The signature of these events would be unique: they would be coincident with galaxies in the nearby Universe,
would not be accompanied by any optical supernovae of the sort that have detected so far, and would often occur
outside of the nuclear regions of the host galaxy.
For especially interesting radio detections, it may be worth conducting followup in the infrared to rule out the presence of an extinguished, radio-bright supernova.
The issue of associated X-ray emission is less certain. 
 It is possible that young magnetars shine in the X-rays, but if the ejecta is dominated by heavy elements then the classical X-ray emission is suppressed.
A background AGN or a foreground active star 
would be revealed by the distinctive spectra of the 
astrometric coincident quiescent source.  

The future is quite bright.
The soon-to-be-commissioned APERTIF \citep{Verheijenetal2008}
mapper could do the same survey 30\% faster or alternatively redo the survey
every few months. In the future ASKAP and MeerKAT mappers could undertake the survey
five times faster.
\\

We thank Roger Chevalier, Peter Goldreich, Gregg Hallinan, Mansi Kasliwal, Keiichi Maeda, Brian Metzger, Christian Ott, E. Sterl Phinney, and Eliot Quataert. ALP was supported through NSF grants AST-1212170, PHY-1151197, and PHY-1068881, NASA ATP grant NNX11AC37G, NSF grant AST-0855535, and the Sherman Fairchild Foundation. SRK's research is in part supported by NSF.

\end{document}